\documentclass[pre,twocolumn,showpacs]{revtex4-1}

\usepackage{amssymb,amsmath}

\usepackage{graphicx}

\newcommand\beq{\begin{equation}}
\newcommand\eeq{\end{equation}}
\newcommand\beqa{\begin{eqnarray}}
\newcommand\eeqa{\end{eqnarray}}
\newcommand{\dd}{\text{d}}

\newcommand{\al}{\alpha}

\usepackage{color}

\begin{document}

\title{Heat flux of driven granular mixtures at low density: Stability analysis of the homogeneous steady state}
\author{Nagi Khalil}
\email{nagi@ifisc.uib-csic.es}
\affiliation{IFISC (CSIC-UIB), Instituto de F\'{\i}sica Interdisciplinar y Sistemas Complejos,
Campus Universitat de les Illes Balears,
E-07122 Palma de Mallorca, Spain}
\author{Vicente Garz\'o}
\email{vicenteg@unex.es} \homepage{http://www.unex.es/eweb/fisteor/vicente/} \affiliation{Departamento de
F\'{\i}sica and Instituto de Computaci\'on Cient\'{\i}fica Avanzada (ICCAEx), Universidad de Extremadura, E-06071 Badajoz, Spain}
\begin{abstract}
The Navier--Stokes order hydrodynamic equations for a low-density driven granular mixture obtained previously [Khalil and Garz\'o, Phys. Rev. E \textbf{88}, 052201 (2013)] from the Chapman--Enskog solution to the Boltzmann equation are considered further. The four transport coefficients associated with the heat flux are obtained in terms of the mass ratio, the size ratio, composition, coefficients of restitution, and the driven parameters of the model. Their quantitative variation on the control parameters of the system is demonstrated by considering the leading terms in a Sonine polynomial expansion to solve the exact integral equations. As an application of these results, the stability of the homogeneous steady state is studied. In contrast to the results obtained in undriven granular mixtures, the stability analysis of the linearized Navier--Stokes hydrodynamic equations shows that the transversal and longitudinal modes are (linearly) stable with respect to long enough wavelength excitations. This conclusion agrees with a previous analysis made for single granular gases.
\end{abstract}


\date{\today}
\maketitle

\section{Introduction}
\label{sec1}

It is well established that when a granular material is externally excited the motion of grains resembles the random motion of atoms or molecules in an ordinary gas. These conditions are referred to as rapid flow conditions and has been an active area of research in the past several decades \cite{G03,BP04,RN08,D09,H13,P15}. On the other hand, since the collisions between grains are inelastic, the energy monotonically decays in time so that external energy must be added to the system to keep it under rapid flow conditions. In real experiments, the external energy is injected into the granular gas from the boundaries (for instance, shearing the system or vibrating its walls \cite{YHCMW02,HYCMW04}), by bulk driving (as in air-fluidized beds \cite{SGS05,AD06}) or by the presence of the interstitial fluid \cite{MLNJ01,YSHLCh03,WZXS08}. When this external energy compensates for the energy dissipated by collisions, then a nonequilibrium steady state is reached.

Nevertheless, when the granular gas is locally driven, strong spatial gradients appear in the bulk domain and consequently the usual Navier--Stokes hydrodynamic equations are not suitable. Thus, in order to avoid the difficulties linked to the theoretical description of far from equilibrium states, it is usual in computer simulations \cite{puglisi,GSVP11,ernst} to drive the gas by means of an external force. Following the terminology employed in nonequilibrium molecular dynamics simulations of ordinary gases \cite{EM90}, this type of external forces are usually called \emph{thermostats}.

Although several kind of thermostats have been proposed in the literature \cite{NE98,MS00}, an interesting and reliable model was proposed by Puglisi and coworkers \cite{puglisi} to homogeneously fluidize a granular gas by an external force. More specifically, the thermostat is constituted by two different terms: (i) a drag force proportional to the velocity of the particle $\mathbf{v}$ and (ii) a stochastic force (Langevin model) where the particles are \emph{randomly} accelerated between successive collisions. This latter force has the form of a Gaussian white force with zero mean and finite variance \cite{WM96}. At a kinetic level and in the low-density regime, the corresponding kinetic equation for this thermostat has the structure of a Fokker-Planck equation \cite{K81,KG14} plus the Boltzmann collision operator. Apart from using these external forces as thermostatic forces, it is worth noting that the above Langevin-like model has the same structure as several kinetic equations proposed in the literature \cite{K90,KH01,GTSH12} to model granular suspensions for low Reynolds numbers. In this context,  the viscous drag force mimics the friction of solid particles on the viscous surrounding fluid while the stochastic force models the transfer of energy from the interstitial fluid to the granular particles.

More recently \cite{KG13}, the Langevin-like model has been extended to the case of granular mixtures. Since the model attempts to incorporate the influence of gas phase into the dynamics of grains, the drag force is defined in terms of the ``peculiar'' velocity $\mathbf{v}-\mathbf{U}_g$ instead of the instantaneous velocity $\mathbf{v}$ of the solid particles. Here, $\mathbf{U}_g$ is the mean velocity of the interstitial gas and is assumed to be a known quantity of the suspension model. Thus, the fluid-solid interaction force is constituted by an additional term (apart from the drag and stochastic forces) proportional to the difference $\mathbf{U}-\mathbf{U}_g$ between the mean velocities of gas and solid phases  ($\mathbf{U}$ being the mean flow velocity of the granular particles).

Aside introducing the model for driven granular binary mixtures at low density, the Navier--Stokes hydrodynamic equations were derived in Ref.\ \cite{KG13} by solving the corresponding Boltzmann kinetic equation by means of the Chapman--Enskog method \cite{CC70}. As in the free cooling case \cite{GD02,GMD06,GDH07,GHD07}, the transport coefficients are given in terms of the solution to a set of coupled linear integral equations. These integral equations can be approximately solved by considering the leading terms in a Sonine polynomial expansion. This task was in part carried out in Ref.\ \cite{KG13} where a complete study of the dependence of the mass flux (four diffusion coefficients) and the shear viscosity coefficient on the parameters of the mixture (masses, sizes, concentration, and coefficients of restitution) was worked out. Therefore, a primary objective here is to demonstrate the variation of the transport coefficients of the heat flux by using the same Sonine polynomial approximation as was found applicable for undriven mixtures \cite{GMD06}. As expected, the results clearly show that the influence of thermostats on heat transport is in general significant since the dependence of inelasticity on the four heat flux transport coefficients is different from the one found before for undriven mixtures \cite{GMD06}.

Needless to say, the knowledge of the complete set of the Navier--Stokes transport coefficients of the mixture opens up the possibility of studying several problems. Among them, an interesting application is to analyze the stability of the homogeneous steady state (HSS). This state plays a similar role to the homogeneous cooling state (HCS) for freely cooling granular flows. It is well known \cite{GZ93,M93} that the HCS becomes unstable when the linear size of the system is larger than a certain critical length $L_c$. The dependence of $L_c$ on the control parameters of the system can be obtained from a linear stability analysis of the Navier--Stokes hydrodynamic equations. The calculation of $L_c$ is interesting by itself and also as a way of assessing the reliability of kinetic theory calculations via a comparison against computer simulations. In fact, previous comparisons made for $L_c$ between theory and simulations for undriven granular gases \cite{BRM98,MDCPH11,MGHEH12,BR13,MGH14} have shown a good agreement even for strong inelasticity and/or disparate masses or sizes in the case of multicomponent systems.

A natural question is whether the HSS may be unstable with respect to long enough wavelength excitations as occurs in the HCS. The results derived here by including the complete dependence of the transport coefficients on the coefficients of restitution show that the HSS is linearly stable with respect to long wavelength perturbations. This conclusion agrees with a previous stability analysis carried out for single driven granular gases \cite{gachve13}. On the other hand, the quantitative forms for the dispersion relations obtained in this paper are quite different from the ones derived for monocomponent inelastic gases.

The plan of the paper is as follows. In Sec.\ \ref{sec2} the model for driven granular mixtures is introduced and the corresponding hydrodynamic equations to Navier--Stokes order are recalled. Next, the four transport coefficients associated with the heat flux are obtained in Sec.\ \ref{sec3} in terms of the parameters of the mixture and the driven parameters of the model. The results for these coefficients are also illustrated for a common coefficient of restitution $\al$ and same size ratio as functions of $\al$ at a concentration $x_1=0.2$ for several values of the mass ratio. The results clearly show a significant deviation of these coefficients from their values for ordinary (elastic) gases, specially for strong inelasticity as expected. Section \ref{sec4} addresses the linear stability analysis around the HSS and present perhaps the main relevant findings of the paper. The paper is closed in Sec.\ \ref{sec5} with some concluding remarks.

\section{Hydrodynamics from Boltzmann kinetic theory}
\label{sec2}

We consider a granular gas modeled as a binary mixture of inelastic hard spheres in $d$ dimensions with masses $m_i$ and diameters $\sigma_i$ ($i=1,2$). The inelasticity of collisions among all pairs is characterized by three independent constant coefficients of normal restitution $\al_{11}$, $\al_{22}$, and $\al_{12}=\al_{21}$, where $0<\al_{ij}\leq 1$. Here, $\al_{ij}$ is the coefficient of restitution for collisions between particles of species $i$ and $j$. We also assume that the particles interact with an external bath. The influence of the bath on the dynamics of grains is encoded in two different terms: (i) a stochastic force assumed to have the form of a Gaussian white noise and (ii) a drag force proportional to the velocity of the particle. Under these conditions, in the low-density regime, the set of coupled nonlinear Boltzmann equations for the one-particle distribution function $f_i(\mathbf{r},\mathbf{v},t)$ of each species reads \cite{KG13}
\beqa
\label{2.1}
\partial_{t}f_i&+&\mathbf{v}\cdot \nabla f_i-\frac{\gamma_\text{b}}{m_i^{\beta}}\Delta \mathbf{U} \cdot
\frac{\partial}{\partial\mathbf{v}}f_i-\frac{\gamma_\text{b}}{m_i^{\beta}}
\frac{\partial}{\partial\mathbf{v}}\cdot \mathbf{V}
f_i\nonumber\\
& & -\frac{1}{2}\frac{\xi_\text{b}^2}{m_i^{\lambda}}\frac{\partial^2}{\partial v^2}f_i=\sum_{j=1}^2\; J_{ij}[f_i,f_j],
\eeqa
where $J_{ij}[f_i,f_j]$ is the Boltzmann collision operator \cite{BP04}. In Eq.\ \eqref{2.1}, $\gamma_b$ is the drag (or friction) coefficient, $\xi_b^2$ is the strength of the correlation in the Gaussian white noise, and $\beta$ and $\lambda$ are arbitrary constants of the driven model. In addition, $\Delta \mathbf{U}=\mathbf{U}-\mathbf{U}_\text{g}$ and $\mathbf{V}(\mathbf{r},t)=\mathbf{v}-\mathbf{U}(\mathbf{r},t)$, where
\beq
\label{2.2}
\mathbf{U}=\rho^{-1}\sum_{i=1}^2\int\dd \mathbf{ v}m_{i}\mathbf{v}f_{i}(\mathbf{ v})
\eeq
is the mean flow velocity of solid particles. In addition, as said before, $\mathbf{U}_g$ can be interpreted as the mean velocity of the fluid surrounding the grains. In Eq.\eqref{2.2}, $\rho=\sum_i m_i n_i$ is the total mass density and
\beq
\label{2.3}
n_{i}=\int \dd\mathbf{ v}\;f_{i}(\mathbf{ v})
\eeq
is the local number density of species $i$. In the case of monodisperse granular gases and for $\beta=1$ and $\lambda=0$, the Boltzmann equation \eqref{2.1} is similar to the one proposed in Ref.\ \cite{GTSH12} to model the effects of the interstitial fluid on grains in monodisperse gas-solid dense suspensions. The only difference between both descriptions is that in the latter case the parameters $\gamma_b$ and $\xi_b^2$ are functions of the Reynolds number, the solid volume fraction, and the difference $\Delta \mathbf{U}$. In this context, the results derived here can be useful for instance to understand the stability of homogeneous bidisperse suspensions.

The parameters $\beta$ and $\lambda$ can be seen as free parameters of the model. Thus, when $\gamma_\text{b}=\lambda=0$, the Boltzmann equation \eqref{2.1} describes the time evolution of a granular mixture driven by the stochastic thermostat employed in several previous works \cite{BT02,DHGD02}. On the other hand, the case $\beta=1$ and $\lambda=2$ reduces to the Fokker--Planck model for ordinary (elastic) mixtures \cite{puglisi}. In this context, our model can be understood as a generalization of previous driven models. Moreover, dimensional analysis clearly shows that the dependence of $\gamma_\text{b}$ and $\xi_\text{b}^2$ on the masses $m_1$ and $m_2$ depends on the specific values of both $\beta$ and $\lambda$ considered in each particular situation.

Apart from the fields $n_i$ and $\mathbf{U}$, the other relevant hydrodynamic field of the mixture is the granular temperature $T(\mathbf{r},t)$. It is defined as
\beq
\label{2.4}
T=\frac{1}{n}\sum_{i=1}^2\int \dd\mathbf{v}\frac{m_{i}}{d}V^{2}f_{i}(\mathbf{v})\;,
\eeq
where $n=n_{1}+n_{2}$ is the total number density. At a kinetic level, it is also convenient to introduce the partial kinetic temperatures $T_i$ for each species defined as
\begin{equation}
\label{2.5}
T_i=\frac{m_{i}}{d n_i}\int\; \dd\mathbf{v}\;V^{2}f_{i}(\mathbf{v}).
\end{equation}
The partial temperatures $T_i$ measure the mean kinetic energy of each species. According to Eq.\ \eqref{2.4}, the granular temperature $T$ of the mixture can also be written as
\beq
\label{2.6}
T=\sum_{i=1}^2\, x_i T_i,
\eeq
where $x_i=n_i/n$ is the concentration or mole fraction of species $i$.

Upon deriving the Boltzmann equation \eqref{2.1}, it has been assumed that the typical collision frequency for collisions between the solid particles and the bath is much larger than the corresponding frequency for particle collisions \cite{KG14}. In addition, given that the Boltzmann collision operator $J_{ij}[f_i,f_j]$ is not affected by the surrounding fluid, one expects then that the suspension model defined by
Eq.\ \eqref{2.1} will accurately describe situations where the stresses exerted by the interstitial fluid on particles are sufficiently
small so they only have a weak influence on the dynamics of grains.

The macroscopic balance equations for the number density of each species $n_i$, flow velocity $\mathbf{U}$, and temperature $T$ can be derived from the set of Boltzmann equations \eqref{2.1}. They are given by  \cite{KG13}
\begin{equation}
D_{t}n_{i}+n_{i}\nabla \cdot \mathbf{U}+\frac{\nabla \cdot \mathbf{ j}_{i}}{m_{i}}
=0,  \label{eq:1}
\end{equation}
\begin{equation}
\label{eq:2}
D_{t}\mathbf{U}+\rho ^{-1}\nabla
\cdot\mathsf{P}=-\frac{\gamma_\text{b}}{\rho}\left(\Delta \mathbf{U}\sum_{i=1}^2\frac{\rho_i}{m_i^\beta}
+\sum_{i=1}^2\frac{\mathbf{ j}_{i}}{m_{i}^\beta}\right),
\end{equation}
\beqa
\label{eq:3}
& & D_{t}T-\frac{T}{n}\sum_{i=1}^2\frac{\nabla \cdot \mathbf{ j}_{i}}{m_{i}}+\frac{2}{dn}
\left( \nabla \cdot \mathbf{ q}+{\sf P}:\nabla \mathbf{U}\right)=-\zeta \,T \nonumber\\
& &
-\frac{2 \gamma_\text{b}}{d n} \sum_{i=1}^2\frac{
\Delta \mathbf{U} \cdot\mathbf{ j}_{i}}{m_{i}^\beta}-2\gamma_\text{b}\sum_{i=1}^2
\frac{x_i T_i}{m_i^\beta}+\frac{\xi_\text{b}^2}{n}\sum_{i=1}^2\frac{\rho_i}{m_i^\lambda}.
\eeqa
In the above equations, $D_t\equiv \partial_t+\mathbf{U}\cdot \nabla$ is the material derivative, $\rho_i=m_i n_i$ is the partial mass density of species $i$, $\mathbf{j}_i$ is the mass flux of species $i$ relative to the local flow, $\mathsf{P}$ is the pressure tensor, $\mathbf{q}$ is the heat flux, and $\zeta$ is the cooling rate. In the case of a binary mixture, there are $d+3$ independent hydrodynamic fields, $n_1$, $n_2$, $\mathbf{U}$, and $T$. To obtain a closed set of hydrodynamic equations, one has to express the fluxes and the cooling rate in terms of the above hydrodynamic fields. These expressions are called ``constitutive equations.'' Such expressions were derived in Ref.\ \cite{KG13} by solving the Boltzmann equation from the Chapman--Enskog method.

It is worth remarking that several physical assumptions are required to obtain the above constitutive equations. First, the hydrodynamics fields $n_i$, $\mathbf{U}$, and $T$ are assumed to be the slowest magnitudes of the system and hence, for any initial condition, all the other magnitudes (for instance, the partial temperatures $T_i$, the fluxes, the cooling rate, $\ldots$) become functionals of the hydrodynamic fields for times longer than the mean free time. Second, the functional dependence of the distribution functions $f_i$ on the hydrodynamic fields is well approximated by the linear terms (Navier--Stokes approximation) of a series expansion in powers of the spatial gradients. The reference state in this expansion is in general supposed to be the local version of the time-dependent homogeneous state described in Refs. \cite{KG14} and \cite{KG13}. Third, the difference of the mean velocities $\Delta \mathbf U$ is assumed to be at least of  first order in the spatial gradients.

Moreover, instead of providing the mass and heat fluxes in terms of the partial densities $n_i$, it is more convenient to express such fluxes in terms of a different set of experimentally more accessible fields like the mole fraction $x_1=n_1/n$ and the pressure $p=n T$. The corresponding hydrodynamic balance equations for $x_1$ and $p$ can be easily derived from Eqs.\ \eqref{eq:1} and \eqref{eq:3} by a simple change of variables. They are given by
\beq
\label{2.7}
D_t x_1+\frac{\rho}{n^2 m_1m_2}\nabla \cdot \mathbf{j}_1=0,
\eeq
\beqa
\label{2.8}
& & D_t p+ p \nabla \cdot \mathbf{U}+\frac{2}{d}\left( \nabla \cdot \mathbf{ q}+\mathsf{P}:\nabla \mathbf{U}\right)=-\zeta \,p
\nonumber\\
& & -\frac{2\gamma_\text{b}}{d}\sum_{i=1}^2
\frac{\Delta \mathbf{U}\cdot \mathbf{j}_i}{m_i^\beta}-2\gamma_\text{b}p\sum_{i=1}^2
\frac{x_i \chi_i}{m_i^\beta}+\xi_\text{b}^2\sum_{i=1}^2\frac{\rho_i}{m_i^\lambda},
\eeqa
where $\chi_i\equiv T_i/T$ is the temperature ratio for species $i$.

The constitutive equations up to the Navier--Stokes order are
\beqa
\label{eq:4}
\mathbf{ j}_{1}&=&-\left(\frac{m_{1}m_{2}n}{\rho }\right) D\nabla x_{1}-
\frac{\rho}{p}D_{p}\nabla p -\frac{\rho}{T}D_T\nabla T\nonumber\\
& &  -D_U \Delta \mathbf{U}, \quad \mathbf{j}_2=-\mathbf{j}_1,
\eeqa
\begin{equation}
\label{eq:5}
\mathbf{q}=-T^{2}D^{\prime \prime}\nabla x_{1}-L\nabla p-\kappa \nabla T-\kappa_U \Delta \mathbf{U},
\end{equation}
\begin{equation}
\label{eq:6}
P_{k\ell}=-\eta \left( \partial_{k}U_{\ell}+\partial_{\ell}U_{k}-\frac{2}{d}\delta_{k\ell}
\nabla \cdot \mathbf{U}\right),
\end{equation}
\beq
\label{eq:6.1}
\zeta=\zeta^{(0)}+\zeta_U \nabla \cdot \mathbf{U}.
\eeq
The transport coefficients in these equations are
\begin{widetext}
\beq
\left(
\begin{array}{c}
D\\
D_p\\
D_T\\
D_U\\
D''\\
L\\
\kappa\\
\kappa_U\\
\eta
\end{array}
\right)=\left(
\begin{array}{c}
\text{diffusion coefficient}\\
\text{pressure diffusion coefficient}\\
\text{thermal diffusion coefficient}\\
\text{velocity diffusion coefficient}\\
\text{Dufour coefficient}\\
\text{pressure energy coefficient}\\
\text{thermal conductivity}\\
\text{velocity conductivity}\\
\text{shear viscosity}
\end{array}
\right).
\eeq

As for elastic collisions \cite{CC70}, the Navier--Stokes transport coefficients are given in terms of the solution to a set of coupled linear integral equations. In the steady state, these integral equations can be approximately solved by considering the leading terms in a Sonine polynomial expansion. The evaluation of the diffusion coefficients $\left\{D, D_p, D_T, D_U \right\}$ and the shear viscosity coefficient $\eta$ was accomplished in Ref. \ \cite{KG13}. The transport coefficients  $\left\{D'', L, \kappa, \kappa_U \right\}$ associated with the heat flux will be explicitly determined in Sec.\ \ref{sec3} in terms of the parameters of the mixture (masses, sizes, concentration, and coefficients of restitution) and the driven parameters $\gamma_b$ and $\xi_b^2$ of the suspension model.

Once the complete set of transport coefficients is known, the Navier--Stokes constitutive equations \eqref{eq:4}--\eqref{eq:6.1} are substituted into the exact balance equations \eqref{eq:2}--\eqref{2.8} to obtain the corresponding Navier--Stokes hydrodynamic equations for a driven binary granular mixture. They are given by
\beq
\label{2.9}
D_{t}x_{1}=\frac{\rho }{n^{2}m_{1}m_{2}}\nabla \cdot \left( \frac{
m_{1}m_{2}n}{\rho}D\nabla x_{1}+\frac{\rho}{p}D_{p}\nabla p  +\frac{\rho}{T}D_T\nabla T+D_U \Delta \mathbf{U}\right) \;,
\eeq
\beqa
\label{2.10}
& & D_{t}U_{\ell }+\rho ^{-1}\nabla _{\ell }p=\rho ^{-1}\nabla _{k}\eta \left(
\nabla_{\ell }U_{k}+\nabla_{k}U_{\ell}-\frac{2}{d}\delta_{k\ell }\nabla\cdot \mathbf{U}\right)+\frac{\gamma_b}{\rho}\frac{m_2^\beta-m_1^\beta}{(m_1m_2)^{\beta}}\nonumber\\
& & \times\left(
\frac{m_{1}m_{2}n}{\rho}D\nabla x_{1}+\frac{\rho}{p}D_{p}\nabla p  +\frac{\rho}{T}D_T\nabla T+D_U \Delta \mathbf{U}\right)-\frac{\gamma_b}{\rho}\frac{\rho_1 m_2^\beta+\rho_2 m_1^\beta}{(m_1 m_2)^\beta}\Delta U_\ell \;,
\eeqa
\beqa
\label{2.11}
& &\left(D_{t}+\zeta^{(0)} \right) T+\frac{2}{d}p\nabla \cdot \mathbf{U}=
-\frac{T}{n}\frac{m_{2}-m_{1}}{m_{1}m_{2}}\nabla \cdot \left(
\frac{m_{1}m_{2}n}{\rho}D\nabla x_{1}+\frac{\rho}{p}D_{p}\nabla p  +\frac{\rho}{T}D_T\nabla T+D_U \Delta \mathbf{U}\right) \nonumber \\
&&+\frac{2}{dn}\nabla \cdot \left( T^{2}D^{\prime \prime }\nabla
x_{1}+L\nabla p+\kappa \nabla T+\kappa_U \Delta \mathbf{U}\right)
+\frac{2}{dn}\eta \left( \nabla_{\ell }U_{k}+\nabla_{k}U_{\ell }-\frac{2
}{d}\delta_{k\ell }\nabla \cdot \mathbf{U}\right) \nabla_{\ell}U_{k}+T \zeta_U \nabla\cdot \mathbf{U}\nonumber \\
&&-2\gamma_\text{b}\sum_{i=1}^2
\frac{x_i T_i}{m_i^\beta}+\frac{\xi_\text{b}^2}{n}\sum_{i=1}^2\frac{\rho_i}{m_i^\lambda}
-\frac{2\gamma_b}{d n}\frac{m_2^\beta-m_1^\beta}{(m_1m_2)^{\beta}}\Delta \mathbf{U}\cdot
\left(
\frac{m_{1}m_{2}n}{\rho}D\nabla x_{1}+\frac{\rho}{p}D_{p}\nabla p  +\frac{\rho}{T}D_T\nabla T+D_U \Delta \mathbf{U}\right),
\eeqa
\begin{eqnarray}
\label{2.12}
& &\left(D_{t}+\zeta^{(0)} \right) p+\frac{d+2}{d}p\nabla \cdot \mathbf{U}=
\frac{2}{d}\nabla \cdot \left( T^{2}D^{\prime \prime }\nabla
x_{1}+L\nabla p+\kappa \nabla T+\kappa_U \Delta \mathbf{U}\right)\nonumber \\
& &
+\frac{2}{d}\eta \left( \nabla_{\ell }U_{k}+\nabla_{k}U_{\ell }-\frac{2
}{d}\delta_{k\ell }\nabla \cdot \mathbf{U}\right) \nabla_{\ell}U_{k}+p\zeta_U \nabla\cdot \mathbf{U}-2\gamma_\text{b}p\sum_{i=1}^2
\frac{x_i \chi_i}{m_i^\beta}+\xi_\text{b}^2\sum_{i=1}^2\frac{\rho_i}{m_i^\lambda}\nonumber \\
& &
-\frac{2\gamma_b}{d}\frac{m_2^\beta-m_1^\beta}{(m_1m_2)^{\beta}}\Delta \mathbf{U}\cdot
\left(
\frac{m_{1}m_{2}n}{\rho}D\nabla x_{1}+\frac{\rho}{p}D_{p}\nabla p  +\frac{\rho}{T}D_T\nabla T+D_U \Delta \mathbf{U}\right).
\end{eqnarray}
\end{widetext}
Here, as mentioned in several previous works \cite{GMD06,G05}, the general form of the cooling rate $\zeta$ should include second-order gradient contributions in Eqs.\ \eqref{2.11} and \eqref{2.12}. However, as shown for a one-component dilute granular gas \cite{BDKS98}, these contributions are found to be very small with respect to the remaining contributions and hence they can be neglected in the evaluation of the cooling rate. Apart from this approximation, Eqs.\ \eqref{2.9}--\eqref{2.12} are exact to second order in the spatial gradients
for a low-density driven granular binary mixture.

\section{Heat flux transport coefficients}
\label{sec3}

The evaluation of the heat flux transport coefficients $D''$, $L$, $\kappa$, and $\kappa_U$ requires to consider the second Sonine approximation. The expressions of these transport coefficients are
\beqa
\label{eq:16}
D''&=&-\frac{d+2}{2}\frac{p}{T(m_1+m_2)\nu_0} \left[\frac{x_1\chi_1^3}{\mu_{12}}d_1^*+\frac{x_2\chi_2^3}{\mu_{21}}d_2^*\right.\nonumber\\
& & \left.-\left(\frac{\chi_1}{\mu_{12}}-\frac{\chi_2}{\mu_{21}}\right) D^* \right],
\eeqa
\beqa
\label{eq:17}
L&=&-\frac{d+2}{2}\frac{T}{(m_1+m_2)\nu_0} \left[\frac{x_1\chi_1^3}{\mu_{12}}\ell_1^*+\frac{x_2\chi_2^3}{\mu_{21}}\ell_2^*\right.\nonumber\\
& & \left.-\left(\frac{\chi_1}{\mu_{12}}-\frac{\chi_2}{\mu_{21}}\right) D_p^* \right],
\eeqa
\beqa
\label{eq:18}
\kappa&=&-\frac{d+2}{2}\frac{p}{(m_1+m_2)\nu_0} \left[\frac{x_1\chi_1^3}{\mu_{12}}\kappa_1^*+\frac{x_2\chi_2^3}{\mu_{21}}\kappa_2^*\right.\nonumber\\
& & \left.-\left(\frac{\chi_1}{\mu_{12}}-\frac{\chi_2}{\mu_{21}}\right) D_T^* \right],
\eeqa
\beqa
\label{eq:19}
\kappa_U&=&-\frac{d+2}{2}\frac{\rho T}{(m_1+m_2)\nu_0} \left[\frac{x_1\chi_1^3}{\mu_{12}}\kappa_{U1}^*+\frac{x_2\chi_2^3}{\mu_{21}}\kappa_{U1}^*\right.\nonumber\\
& & \left.-\left(\frac{\chi_1}{\mu_{12}}-\frac{\chi_2}{\mu_{21}}\right) D_U^* \right],
\eeqa
where $\mu_{ij}=m_i/(m_i+m_j)$. In Eqs.\ \eqref{eq:16}--\eqref{eq:19}, the explicit forms of the dimensionless Sonine coefficients $d_i^*$, $\ell_i^*$, $\kappa_i^*$, and $\kappa_{Ui}^*$ have been determined in the Appendix. Moreover, the (reduced) diffusion coefficients $D^*$, $D_p^*$, $D_T^*$, and $D_U^*$ are defined by the relations
\begin{equation}
\label{eq:9}
\begin{split}
D=\frac{\rho T}{m_{1}m_{2}\nu_{0}}D^*,\quad    D_{p}=\frac{nT}{\rho\nu_{0}}D_{p}^{\ast},\\
D_T=\frac{nT}{\rho \nu_{0}}D_T^*, \quad D_U=\frac{p\overline m}{2T} D_U^*,
\end{split}
\end{equation}
where
\beq
\label{overlinem}
\overline{m}=\frac{m_1m_2}{m_1+m_2}
\eeq
is the reduced mass and
\beq
\label{nu0}
\nu_0=\frac{p}{T}\sigma_{12}^{d-1}v_0
\eeq
is an effective collision frequency. Here, $v_0=\sqrt{2T/\overline{m}}$ is a thermal speed for a binary mixture and $\sigma_{12}=(\sigma_1+\sigma_2)/2$. The above diffusion coefficients were obtained in Ref.\ \cite{KG13} in the first Sonine approximation.

Before considering a binary mixture, it is interesting to check the consistency between the present expression of $\mathbf{q}$ with the one derived for a monocomponent granular gas in Ref.\ \cite{gachve13} when $\Delta \mathbf{U}=\mathbf{0}$. Therefore, for mechanically equivalent particles ($m_1=m_2\equiv m$, $\sigma_1=\sigma_2\equiv \sigma$, $\al_{ij}\equiv \al$), the Dufour coefficient $D''$ vanishes as expected and the heat flux \eqref{eq:5} can be written as
\beq
\label{vic1}
\mathbf{q}=-\overline{\kappa}\nabla T-\overline{\mu} \nabla n,
\eeq
where
\beq
\label{vic2}
\overline{\kappa}=\kappa+n L, \quad \overline{\mu}=T L.
\eeq
Note that the relation $\nabla p=n\nabla T+T \nabla n$ has been used in writing the heat flux $\mathbf{q}$ in the form \eqref{vic1}. In the limit of mechanically equivalent particles, the results derived in the Appendix yield
\beq
\label{vic3}
\overline{\kappa}=\frac{d+2}{2}\frac{p}{m \nu_0} \left(\nu_{11}+\nu_{12}+2^{-\lambda}\xi^*-2\zeta^*\right)^{-1},
\eeq
\beq
\label{vic4}
\overline{\mu}=\frac{\zeta^*}{\nu_{11}+\nu_{12}+\frac{3}{2^\beta}\omega^*\xi^{*1/3}}\frac{T \overline{\kappa}}{n},
\eeq
where $\zeta^*\equiv \zeta^{(0)}/\nu_0$ and the remaining quantities are defined in the Appendix. Equations \eqref{vic3} and \eqref{vic4} agree with the expressions obtained in Ref.\ \cite{gachve13} when $\lambda=0$ and $\beta=1$ in the case that one neglects non-Gaussian corrections to the zeroth-order distribution function (namely, by taking $a_2=0$ in the expressions provided in Ref.\ \cite{gachve13}). This confirms the relevant known limiting case for the granular mixture results described here.

Furthermore, in order to compare the present results with those obtained for \emph{undriven} granular mixtures \cite{GD02,GMD06}, it is convenient to consider an equivalent representation in terms of the heat flow $\mathbf{J}_q$ defined as \cite{GM84}
\begin{equation}
\label{Jq}
\mathbf J_{q}\equiv \mathbf{q}-\frac{d+2}{2}T\sum_{i=1}^2\; \frac{\mathbf{j}_i}{m_i}=\mathbf q -\frac{d+2}{2}T\frac{m_2-m_1}{m_1m_2}\mathbf j_1,
\end{equation}
where use has been made of the requirement $\mathbf{j}_2=-\mathbf{j}_1$ in the second equality of Eq.\ \eqref{Jq}. The difference between $\mathbf{q}$ and $\mathbf{J}_q$ is the contribution to the heat flux coming from the diffusion flux $\mathbf{j}_1$. In particular, for ordinary mixtures (elastic collisions) and in the context of Onsager's reciprocal relations \cite{GM84}, $\mathbf{J}_q$ is the flux conjugate to the temperature gradient in the form of the entropy production. Moreover, the thermal conductivity coefficient in a mixture is measured in the absence of diffusion (i.e., when $\mathbf{j}_1=\mathbf{0}$). To identify this coefficient one has to express $\mathbf{J}_q$ in terms of $\mathbf{j}_1$, $\nabla T$, $\nabla p$, and $\Delta \mathbf{U}$. In this representation, the corresponding coefficient of $\nabla T$ defines the thermal conductivity coefficient \cite{GM84}. To express $\mathbf{J}_q$ in terms of $\mathbf{j}_1$, one may use Eq.\ \eqref{eq:4} to write the gradient of mole fraction $\nabla x_1$ as
\beqa
\label{nablax1}
\nabla x_1&=&-\frac{\rho}{m_1m_2nD}\mathbf{j}_1-\frac{\rho^2 D_p}{m_1m_2 n p D}\nabla p-\frac{\rho^2 D_T}{m_1m_2 p D}\nabla T
\nonumber\\
& & -\frac{\rho D_U}{m_1m_2nD}\Delta \mathbf{U}.
\eeqa
The heat flow $\mathbf{J}_q$ is obtained by substituting Eq.\ \eqref{eq:5} into Eq.\ \eqref{Jq} and eliminating $\nabla x_1$ by using the relation \eqref{nablax1}. The final form of $\mathbf{J}_q$ is
\begin{equation}
\label{Jq.1}
\mathbf J_{q}=\frac{p\rho}{m_1m_2 n^2} \kappa_T \mathbf j_1-\kappa'\mathbf \nabla T-L_p \nabla p-\kappa_U' \Delta \mathbf U,
\end{equation}
where
\beq
\label{Jq.2}
\kappa_T=\frac{T D''}{D}-\frac{d+2}{2}\frac{n}{\rho}(m_2-m_1),
\eeq
\beq
\label{Jq.3}
\kappa'=\kappa-\frac{\rho^2 T D'' D_T}{m_1m_2nD},
\eeq
\beq
\label{Jq.4}
L_p=L-\frac{\rho^2 T D''D_p}{m_1m_2n^2D},
\eeq
\beq
\label{Jq.5}
\kappa_U'=\kappa_U-\frac{\rho T^2 D''D_U}{m_1m_2nD}.
\eeq
As for elastic collisions, the coefficient $\kappa'$ is the thermal conductivity while $\kappa_T$ is called the thermal diffusion (Soret) factor. The coefficient $\kappa_U'$ is also present for ordinary mixtures (if both species are mechanically different) while there is a new contribution to $\mathbf{J}_q$ proportional to $\nabla p$ that vanishes for elastic collisions. This latter contribution defines the transport coefficient $L_p$.
\begin{figure}[!h]
  \centering
  \includegraphics[width=.45\textwidth]{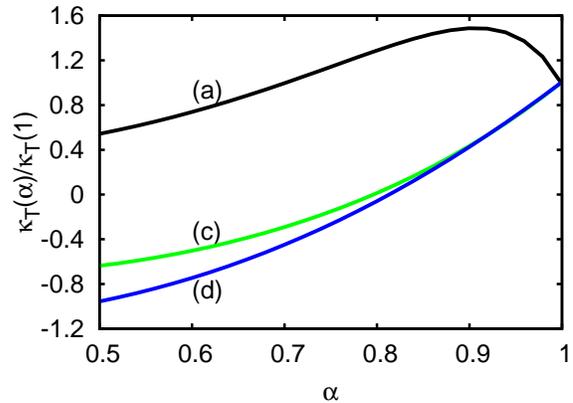}
  \caption{Plot of the (reduced) thermal diffusion factor $\kappa_T(\al)/\kappa_T(1)$ as a function of the (common) coefficient of restitution for a three-dimensional granular binary mixture with $x_1=0.2$, $\sigma_1/\sigma_2=1$, and three different values of the mass ratio  $m_1/m_2$: $m_1/m_2=0.5$ (a), $m_1/m_2=2$ (c), and $m_1/m_2=4$ (d). The parameters of the driven model are $\gamma_\text{b}=0.1$, $\xi_\text{b}^2=0.2$, $\lambda=2$, and $\beta=1$.}
  \label{fig:1}
\end{figure}

\begin{figure}[!h]
  \centering
  \includegraphics[width=.45\textwidth]{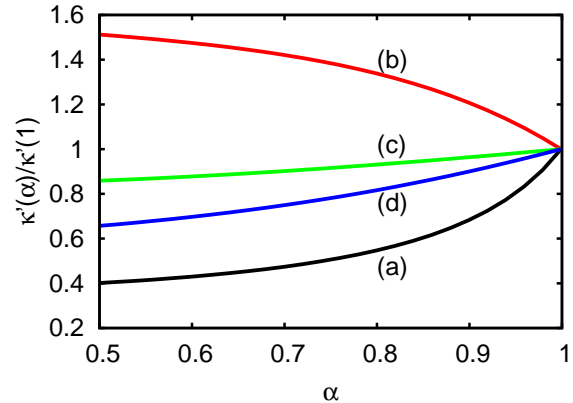}
  \caption{Plot of the (reduced) thermal conductivity coefficient $\kappa'(\al)/\kappa'(1)$ as a function of the (common) coefficient of restitution for a three-dimensional granular binary mixture with $x_1=0.2$, $\sigma_1/\sigma_2=1$, and four different values of the mass ratio  $m_1/m_2$: $m_1/m_2=0.5$ (a), $m_1/m_2=1$ (b), $m_1/m_2=2$ (c), and $m_1/m_2=4$ (d). The parameters of the driven model are $\gamma_\text{b}=0.1$, $\xi_\text{b}^2=0.2$, $\lambda=2$, and $\beta=1$.}
  \label{fig:2}
\end{figure}

\begin{figure}[!h]
  \centering
  \includegraphics[width=.45\textwidth]{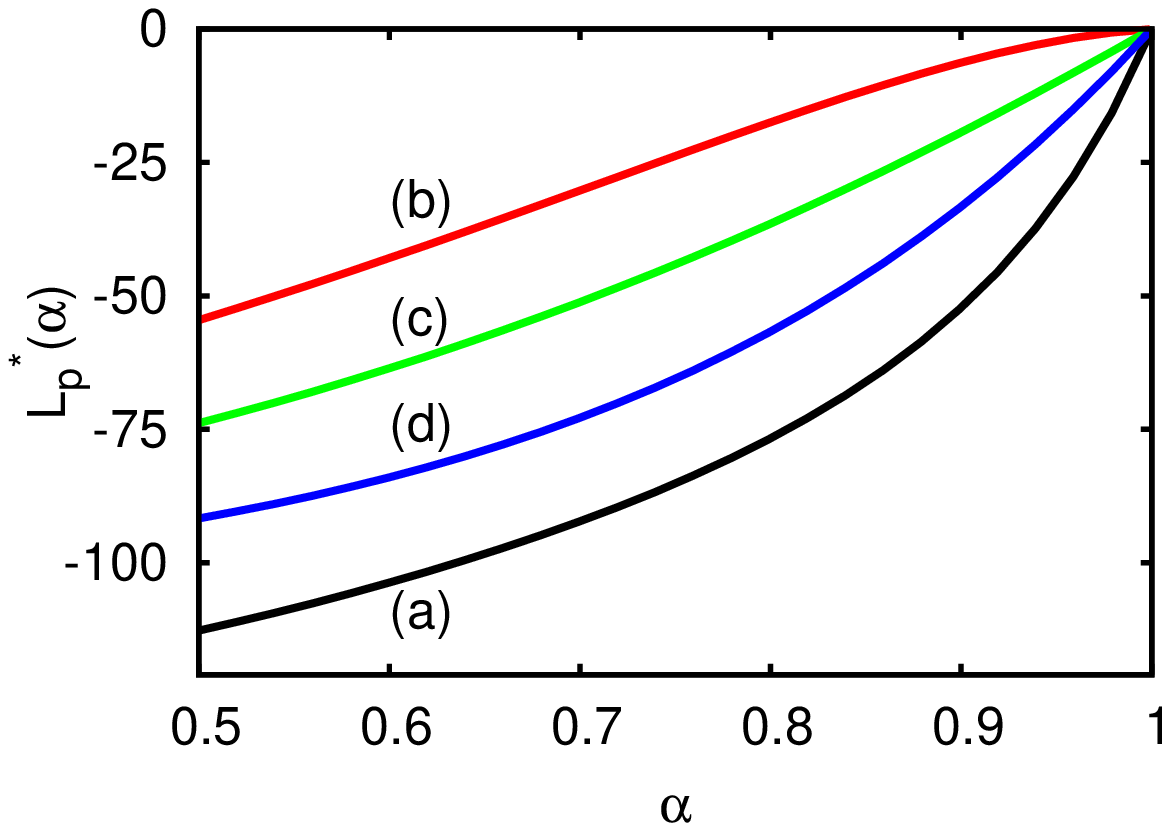}
  \caption{Plot of the (reduced) transport coefficient $L_p^*(\al)$ as a function of the (common) coefficient of restitution for a three-dimensional granular binary mixture with $x_1=0.2$, $\sigma_1/\sigma_2=1$, and four different values of the mass ratio  $m_1/m_2$: $m_1/m_2=0.5$ (a), $m_1/m_2=1$ (b), $m_1/m_2=2$ (c), and $m_1/m_2=4$ (d). The parameters of the driven model are $\gamma_\text{b}=0.1$, $\xi_\text{b}^2=0.2$, $\lambda=2$, and $\beta=1$. Note that $L_p^*$ vanishes for elastic collisions.}
  \label{fig:3}
\end{figure}

\begin{figure}[!h]
  \centering
  \includegraphics[width=.45\textwidth]{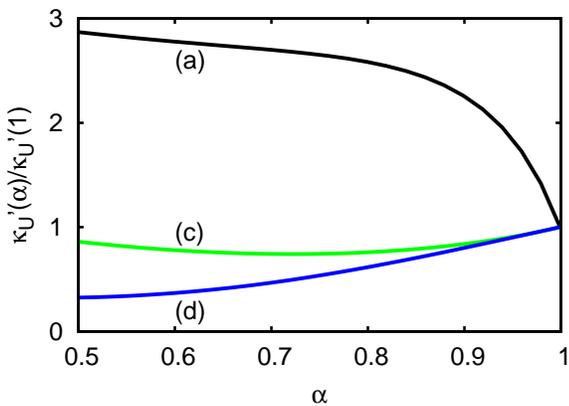}
  \caption{Plot of the (reduced) transport coefficient $\kappa_U'(\al)/\kappa_U'(1)$ as a function of the (common) coefficient of restitution for a three-dimensional granular binary mixture with $x_1=0.2$, $\sigma_1/\sigma_2=1$, and three different values of the mass ratio  $m_1/m_2$: $m_1/m_2=0.5$ (a), $m_1/m_2=2$ (c), and $m_1/m_2=4$ (d). The parameters of the driven model are $\gamma_\text{b}=0.1$, $\xi_\text{b}^2=0.2$, $\lambda=2$, and $\beta=1$. Note that $\kappa_U'$ vanishes for mechanically equivalent particles.}
  \label{fig:4}
\end{figure}

\subsection{Some illustrative systems}

It is quite apparent that the dimensionless forms of the heat flux transport coefficients depend on many parameters: $\left\{x_1, \sigma_1/\sigma_2, m_1/m_2, \al_{11}, \al_{22}, \al_{12}\right\}$. Moreover, the driven parameters ($\gamma_\text{b}$ and $\xi_\text{b}^2$) along with the parameters $\beta$ and $\lambda$ characterizing the class of model considered must be also specified. A complete exploration of the full parameters space is simple but beyond the goal of this paper. Thus, to illustrate the differences between ordinary and granular mixtures the transport coefficients are scaled with respect to their values in the elastic limit. In addition, for the sake of simplicity, we consider a common coefficient of restitution ($\al_{11}=\al_{22}=\al_{12}\equiv \al$), a common size $\sigma_1=\sigma_2$, a mole fraction $x_1=0.2$, and four different values of the mass ratio: $m_1/m_2=0.5, 1, 2,$ and $4$. These are the same systems as those studied before for undriven granular mixtures \cite{GMD06}. We choose a three-dimensional system ($d=3$) with $\gamma_\text{b}=0.1$, $\xi_\text{b}^2=0.2$, $\lambda=2$, and $\beta=1$.

Figures \ref{fig:1}--\ref{fig:4} show the heat flux transport coefficients versus the (common) coefficient of restitution $\al$ for the systems mentioned before. It is understood that all the coefficients have been evaluated with respect to their elastic values, except the case of $L_p$ since this coefficient vanishes for elastic collisions. In this latter case, we have considered the (reduced) coefficient
\beq
\label{L_p^*}
L_p^*=-\frac{m_1+m_2}{\frac{d+2}{2}\sqrt{\pi} T \nu_0}L_p.
\eeq
Figure \ref{fig:1} shows the thermal diffusion factor $\kappa_T$. Note that $\kappa_T=0$ when $m_1=m_2$. It is seen that while $\kappa_T$ exhibits a non-monotonic dependence on inelasticity when the defect component $1$ is lighter than the excess component $2$, this transport coefficient decreases monotonically with decreasing $\al$ when $m_1>m_2$. Moreover, the impact of inelasticity on the functional form of thermal diffusion is more important for $m_1>m_2$ than in the opposite case. The thermal conductivity is shown in Fig.\ \ref{fig:2}. We observe that the (scaled) thermal conductivity decreases with increasing inelasticity, except for mechanically equivalent particles ($m_1=m_2$). This behavior contrasts clearly with the results found for undriven mixtures \cite{GMD06} where the ratio $\kappa'(\al)/\kappa'(1)$ increases with increasing dissipation, regardless of the value of the mass ratio. It is also apparent that, at a given value of $\al$, there is a significant mass dependence of the thermal conductivity, especially at strong inelasticity. The coefficient $L_p^*$ is illustrated in Fig.\ \ref{fig:3} where it is found that its magnitude is larger than the one obtained before for both the thermal diffusion and thermal conductivity coefficients. As for undriven mixtures \cite{GMD06}, $L_p^*$ is negative for the cases studied here. Finally, the (scaled) transport coefficient $\kappa_U'$ is plotted in Fig.\ \ref{fig:4}. This coefficient vanishes for mechanically equivalent particles. In contrast to the thermal conductivity coefficient, $\kappa_U'$ increases with inelasticity when $m_1<m_2$ while the opposite happens when $m_1>m_2$. Furthermore, the impact of collisional dissipation on this coefficient is quite significant for small mass ratios.

In summary, the heat flux transport coefficients for driven granular mixtures differ noticeably from those for ordinary mixtures, especially at strong inelasticity. Depending on the value of the mass ratio, in some cases these transport coefficients decrease with decreasing $\al$ while in others they increase with inelasticity. This non-monotonic behavior with the mass ratio is also present in the free cooling case \cite{GMD06}. Since the expressions of the heat flux transport coefficients obtained here are quite complex (in contrast to more phenomenological approaches), it is very difficult to provide a simple explanation of the non-monotonic trend in the mass ratio observed in Figs.\  \ref{fig:1}--\ref{fig:4}. Finally, with respect to the impact of masses on the heat transport, we also observe that in general the transport coefficients $\kappa_T$, $\kappa'$, $L_p^*$, and $\kappa_U'$ exhibit a strong influence of the mass ratio, being this influence larger than the one found in undriven granular mixtures \cite{GMD06}.


\section{Linear stability analysis of the homogeneous steady state}
\label{sec4}

It is well known for undriven granular mixtures that the HCS is unstable against long enough wavelength perturbations \cite{GMD06,G15}. These instabilities can be well predicted by a linear stability analysis of the Navier--Stokes hydrodynamic equations. In fact, the solution of the linearized hydrodynamic equations provides a critical length $L_c$ beyond which the system becomes unstable. The theoretical predictions of $L_c$ \cite{G05,GMD06,G15} have been shown to compare quite well with computer simulations for monocomponent \cite{BRM98,MDCPH11,MGHEH12} and multicomponent \cite{BR13,MGH14} granular fluids.

On the other hand, a similar analysis for single driven granular fluids \cite{gachve13} has clearly shown that the HSS is always (linearly) stable. This analytical finding agrees with the results obtained from Langevin dynamics simulations \cite{GSVP11}. An interesting question arises then as to whether, and if so to what extent, the conclusions drawn for monocomponent driven granular gases \cite{GSVP11,gachve13} may be altered when a binary mixture is considered.

In order to analyze the stability of the homogeneous solution, Eqs.\ \eqref{2.9}--\eqref{2.11} must be linearized around the HSS. In this state, the hydrodynamic fields take the steady values $x_{1s}= \text{const.}$, $p_s= \text{const.}$, $T_s= \text{const.}$, and $\mathbf{U}_s=\mathbf{U}_{gs}=\mathbf{0}$, where the subscript $s$ denotes the steady state. Moreover, in reduced units, the steady state condition determining the temperature ratios $\chi_{is}$ is
\beq
\label{4.0}
2 M_i^{-\beta}\chi_{is} \gamma_s^*+\zeta_{is}^* \chi_{is}=M_{i}^{1-\lambda} \xi_s^*, \quad i=1,2
\eeq
where $M_i\equiv m_i/\overline{m}$, $\zeta_{is}^*\equiv \zeta_{is}^{(0)}/\nu_{0s}$ is the (reduced) partial cooling rate, and
\beq
\label{4.0.1}
\gamma_s^*\equiv \frac{\gamma_{\text{b}}}{\overline{m}^\beta \nu_{0s}}, \quad \xi_s^*\equiv \frac{\xi_{\text{b}}^2}{\overline{m}^{\lambda-1} T_s \nu_{0s}}.
\eeq
Here, $\nu_{0s}$ is defined by Eq.\ \eqref{nu0} with the replacements $p\to p_s$ and $T\to T_s$. It is interesting to note that for elastic collisions and mechanically different particles energy equipartition is fulfilled ($\chi_{is}=1$) only when $\lambda=\beta+1$ \cite{KG13}. This is the expected result in agreement with the fluctuation-dissipation relation \cite{KTH85}. The above relation between $\lambda$ and $\beta$ will be kept in the remaining part of this section.

We assume that the deviations $\delta y_\mu(\mathbf{r},t)=y_\mu(\mathbf{r},t)-y_{\mu s}$ are small where $\delta y_\mu$ denotes
the deviations of $x_1$, $p$, $T$, and $\mathbf{U}$ from their values in the HSS. Moreover, as usual we also suppose that the interstitial fluid is not perturbed and hence, $\mathbf{U}_g=\mathbf{U}_{gs}=\mathbf{0}$.

In order to compare the present results with those obtained for undriven granular mixtures \cite{GMD06}, the following dimensionless space and time variables are introduced:
\beq
\label{4.1}
\dd \tau=\nu_{0s} \dd t, \quad \dd \mathbf{r}'=\frac{\nu_{0s}}{v_{0s}}\dd \mathbf{r},
\eeq
where $v_{0s}=\sqrt{2T_s/\overline{m}}$. Then, as usual, a set of Fourier transform dimensionless variables are defined as
\beq
\label{4.2}
\rho_{1{\bf k}}(\tau)=\frac{\delta x_{1{\bf k}}(\tau)}{x_{1s}}, \quad
{\bf w}_{{\bf k}}(\tau)=\frac{\delta {\bf U}_{{\bf k}}(\tau)}{v_{0s}},
\eeq
\beq
\label{4.3}
\quad \theta_{{\bf k}}(\tau)=\frac{\delta T_{{\bf k}}(\tau)}{T_{s}}, \quad
\Pi_{{\bf k}}(\tau)=\frac{\delta p_{{\bf k}}(\tau)}{p_{s}},
\end{equation}
where $\delta y_{{\bf k}\mu}\equiv \{\delta \rho_{1{\bf k}}(\tau),
{\bf w}_{{\bf k}}(\tau), \theta_{{\bf k}}(\tau), \Pi_{{\bf k}}(\tau)\}$ is defined as
\begin{equation}
\label{4.45}
\delta y_{{\bf k}\mu}(\tau)=\int \dd \mathbf{r}'\;
e^{-i{\bf k}\cdot \mathbf{r}'}\delta y_{\mu}(\mathbf{r}',\tau).
\end{equation}
Note that in Eq.\ \eqref{4.45} the wave vector ${\bf k}$ is dimensionless.

In terms of the above dimensionless variables, as usual, the $d-1$ transverse velocity components ${\bf w}_{{\bf k}\perp}={\bf w}_{{\bf k}}-({\bf w}_{{\bf k}}\cdot \widehat{{\bf k}})\widehat{{\bf k}}$ (orthogonal to the wave vector ${\bf k}$)
decouple from the other four modes and hence can be obtained more
easily. Their evolution equation is
\beq
\label{4.5}
\frac{\partial {\bf w}_{{\bf k}\perp}}{\partial \tau}=\lambda_\perp {\bf w}_{{\bf k}\perp},
\eeq
where the eigenvalue $\lambda_\perp$ is given by
\beq
\label{4.6}
\lambda_\perp=-\frac{\overline{m}p}{2\rho T}\eta^* k^2+\gamma^*
\left[\frac{\overline{m}p}{2\rho T}\delta m_\beta D_U^*-\frac{\rho_1 m_2^\beta+\rho_2 m_1^\beta}{\rho(m_1+m_2)^\beta}\right],
\eeq
where
\beq
\label{a43}
\delta m_\beta=\frac{m_2^\beta-m_1^\beta}{(m_1+m_2)^\beta},
\eeq
and $\eta^*\equiv\ (\nu_0/p)\eta$. Note that although the subscript $s$ has been omitted in Eqs.\ \eqref{4.5}--\eqref{a43} for the sake of brevity, it is understood that all the quantities are evaluated in the HSS.

The solution to Eq.\ \eqref{4.5} is simply given by
\beq
\label{4.7}
{\bf w}_{{\bf k}\perp}(\mathbf{k},\tau)={\bf w}_{{\bf k}\perp}(0)e^{\lambda_\perp \tau}.
\eeq
For mechanically equivalent particles, $D_U^*=0$ and so, $\lambda_\perp < 0$. This means that the transversal shear mode is always stable for monocomponent granular gases. This finding is consistent with the results derived for simple granular fluids \cite{gachve13}.
In the general case, the (reduced) transport coefficient $D_U^*$ is \cite{KG13}
\beq
\label{4.7.1}
D_U^*=\frac{2T}{\rho {\overline m} p}\frac{\omega^* \xi^{*1/3}}{a_{11}}\delta m_\beta,
\eeq
where
\beq
\label{4.7.2}
a_{11}=\nu_D+\frac{\rho_1 m_2^\beta+\rho_2 m_1^\beta}{\rho(m_1+m_2)^\beta} \omega^* \xi^{*1/3},
\eeq
and
\beqa
\label{4.7.3}
\nu_D&=&\frac{2\pi^{(d-1)/2}}{d\Gamma\left(\frac{d}{2}\right)}(1+\al_{12})\left(\frac{M_1\chi_2+M_2\chi_1}{M_1 M_2}\right)^{1/2}
\nonumber\\
& & \times
\left(x_2 M_1^{-1}+x_1 M_2^{-1}\right).
\eeqa
Here, $\omega^*$ is defined by Eq.\ \eqref{a15}. Since $\nu_D >0$, according to Eqs.\ \eqref{4.7.1} and \eqref{4.7.2} it is easy to prove that
\beq
\label{4.7.45}
D_U^* \leq \frac{2T}{p\overline m}\rho_1\rho_2\frac{m_2^\beta-m_1^\beta}{\rho_1m_1^\beta+\rho_2m_2^\beta},
\eeq
and so, one infers the general result
\begin{equation}
\label{4.8}
\lambda_\perp\le -\frac{\overline{m}p}{2\rho T}\eta^* k^2-\gamma^*\frac{\rho\overline m^\beta}{\rho_1m_1^\beta+\rho_2m_2^\beta}<0.
\end{equation}
Therefore, the transversal shear mode ${\bf w}_{{\bf k}\perp}$ is always (linearly) stable.

The remaining (longitudinal) four modes are the concentration field $\rho_{1\mathbf{k}}$, the temperature field $\theta_{\mathbf{k}}$, the pressure field $\Pi_{\mathbf{k}}$, and the longitudinal component of the velocity field ${\bf w}_{{\bf k}\parallel}={\bf w}_{{\bf k}}\cdot \widehat{{\bf k}}$ (parallel to ${\bf k}$). As in the undriven case \cite{GMD06}, they are coupled and obey the time-dependent equation
\beq
\label{4.9}
\frac{\partial \delta z_{\mathbf{k}\mu} (\tau)}{\partial \tau}=\left(M_{\mu \nu}^{(0)}+i k M_{\mu \nu}^{(1)}
+k^2 M_{\mu \nu}^{(2)}\right)\delta z_{\mathbf{k}\nu}(\tau),
\eeq
where here $\delta z_{\mathbf{k}\mu}$ denotes the set of four variables $(\rho_{1\mathbf{k}}, \theta_{\mathbf{k}}, \Pi_{\mathbf{k}}, \mathbf{w}_{\mathbf{k}\parallel})$. The square matrices in Eq.\ \eqref{4.9} are
\beq
\label{4.10}
M_{\mu \nu}^{(0)}=-\zeta^* A_{\mu \nu}-2\gamma^* B_{\mu \nu}+\xi^* C_{\mu \nu},
\eeq
\beq
\label{4.11}
A_{\mu \nu}=\left(
\begin{array}{cccc}
0&0&0&0\\
\zeta_{x_1}&1+\zeta_{T}&\zeta_{p}&0\\
\zeta_{x_1}&\zeta_{T}&1+\zeta_{p}&0\\
0&0&0&0
\end{array}
\right),
\eeq
\begin{widetext}
\beq
\label{4.12}
B_{\mu \nu}=\left(
\begin{array}{cccc}
0&0&0&0\\
\delta m_\beta \chi_{1,x_1}&\mu_{12}^\beta+\delta m_\beta x_1(\chi_1+\chi_{1,T})&\delta m_\beta x_1 \chi_{1,p}&0\\
\delta m_\beta \chi_{1,x_1}&\delta m_\beta x_1 \chi_{1,T}&\mu_{12}^\beta+\delta m_\beta x_1 (\chi_1+ \chi_{1,p})&0\\
0&0&0&-\frac{\overline{m}p}{4\rho T}\delta m_\beta D_U^*+\frac{\rho_1 m_2^\beta+\rho_2 m_1^\beta}{2\rho(m_1+m_2)^\beta}
\end{array}
\right),
\eeq
\beq
\label{4.12}
C_{\mu \nu}=\left(
\begin{array}{cccc}
0&0&0&0\\
x_1 \delta m_{\lambda-1}&0&0&0\\
x_1 \delta m_{\lambda-1}&-(\mu_{12}^{\lambda-1}+x_1 \delta m_{\lambda-1})&\mu_{12}^{\lambda-1}+x_1 \delta m_{\lambda-1}&0\\
0&0&0&0
\end{array}
\right),
\eeq
\beq
\label{4.13}
M_{\mu \nu}^{(1)}=\left(
\begin{array}{cccc}
0&0&0&\frac{\rho \overline{m}}{2n m_1 m_2 x_1}D_U^*\\
0&0&0&-\frac{2}{d}-\frac{1}{2}\delta m_1 D_U^*+\frac{\kappa_U^*}{d}\\
0&0&0&-\frac{d+2}{d}+\frac{\kappa_U^*}{d}\\
\frac{x_1 p}{2\rho T}\overline{m}\delta m_\beta D^* \gamma^* &\frac{p}{2\rho T}\overline{m}\delta m_\beta D_T^* \gamma^* &\frac{p}{2\rho T}\overline{m}\left(\delta m_\beta \gamma^* D_p^*-1\right)&0
\end{array}
\right),
\eeq
\beq
\label{4.14}
M_{\mu \nu}^{(2)}=\left(
\begin{array}{cccc}
-\frac{\overline{m}\rho}{2 n m_1 m_2}D^*&-\frac{\overline{m}\rho}{2 n_1 m_1 m_2}D_T^*&-\frac{\overline{m}\rho}{2 n_1 m_1 m_2}D_p^*&0\\
-x_1\left(\frac{1}{d}D''^{*}-\frac{1}{2}\delta m_1 D^*\right)&-\frac{1}{d}\kappa^*+\frac{1}{2}\delta m_1 D_T^*&-\frac{1}{d}L^*+\frac{1}{2}\delta m_1 D_p^*&0\\
-\frac{x_1}{d}D''^{*}&-\frac{\kappa^*}{d}&-\frac{L^*}{d}&0\\
0&0&0&-\frac{d-1}{d}\frac{\overline{m}p}{\rho T}\eta^*
\end{array}
\right).
\eeq
\end{widetext}
In Eqs.\ \eqref{4.11} and \eqref{4.12}, we have introduced the shorthand notations
\beq
\label{4.15}
\zeta_{x_1}\equiv x_1 \left(\frac{\partial \ln \zeta^{(0)}}{\partial x_1}\right)_{p,T}, \quad \zeta_{T}\equiv T
\left(\frac{\partial \ln \zeta^{(0)}}{\partial T}\right)_{x_1,p},
\eeq
\beq
\label{4.15.1}
\zeta_{p}\equiv p \left(\frac{\partial \ln \zeta^{(0)}}{\partial p}\right)_{x_1,T}, \quad
\chi_{1,x_1}\equiv x_1 \left(\frac{\partial}{\partial x_1}(x_1 \chi_1)\right)_{p,T},
\eeq
\beq
\label{4.16}
\chi_{1,T}\equiv T \left(\frac{\partial \chi_1}{\partial T}\right)_{x_1,p},
\quad \chi_{1,p}\equiv p \left(\frac{\partial \chi_1}{\partial p}\right)_{x_1,T}.
\eeq
In addition, the following set of reduced transport coefficients have been defined:
\beq
\label{4.17}
D''^*\equiv \frac{T\overline m \nu_0}{p}D'', \quad \kappa^*\equiv \frac{\overline m \nu_0}{p} \kappa,
\eeq
\beq
\label{4.18}
L^*\equiv \frac{\overline m \nu_0}{T}L,\quad \kappa_U^* \equiv \frac{1}{p}\kappa_U.
\eeq
As before, the subscript $s$ has been omitted in Eqs.\ \eqref{4.10}--\eqref{4.14} for the sake of brevity. The derivatives of $\zeta_0$ and $\chi_1$ with respect to $x_1$, $T$, and $p$ have been evaluated in Ref.\ \cite{KG14}. In addition, the first-order contribution $\zeta_U$ to the cooling rate has been neglected in Eq.\ \eqref{4.13} since its magnitude is in general very small.

The longitudinal four modes have the form $\exp[\lambda_n(k)\tau]$ for $n=1,2,3,4$, where $\lambda_n(k)$ are the eigenvalues of the square matrix
\beq
\label{4.19}
M_{\mu \nu}\equiv M_{\mu \nu}^{(0)}+i k M_{\mu \nu}^{(1)}+k^2 M_{\mu \nu}^{(2)}.
\eeq
This implies that the eigenvalues $\lambda$ are the solutions of the quartic equation
\beq
\label{4.20}
X(k,\lambda)\equiv \det \left(\mathsf{M}-\lambda \openone \right)=0,
\eeq
where $\openone$ is the identity matrix. It is quite apparent that the dependence of the longitudinal modes on the wave vector $k$ is quite intricate. Thus, in order to gain some insight into the general problem, it is instructive first to analyze the solutions to the hydrodynamic equations in the extreme long wavelength limit ($k=0$ or Euler hydrodynamic equations).

\begin{figure}[!h]
  \centering
  \includegraphics[width=.45\textwidth]{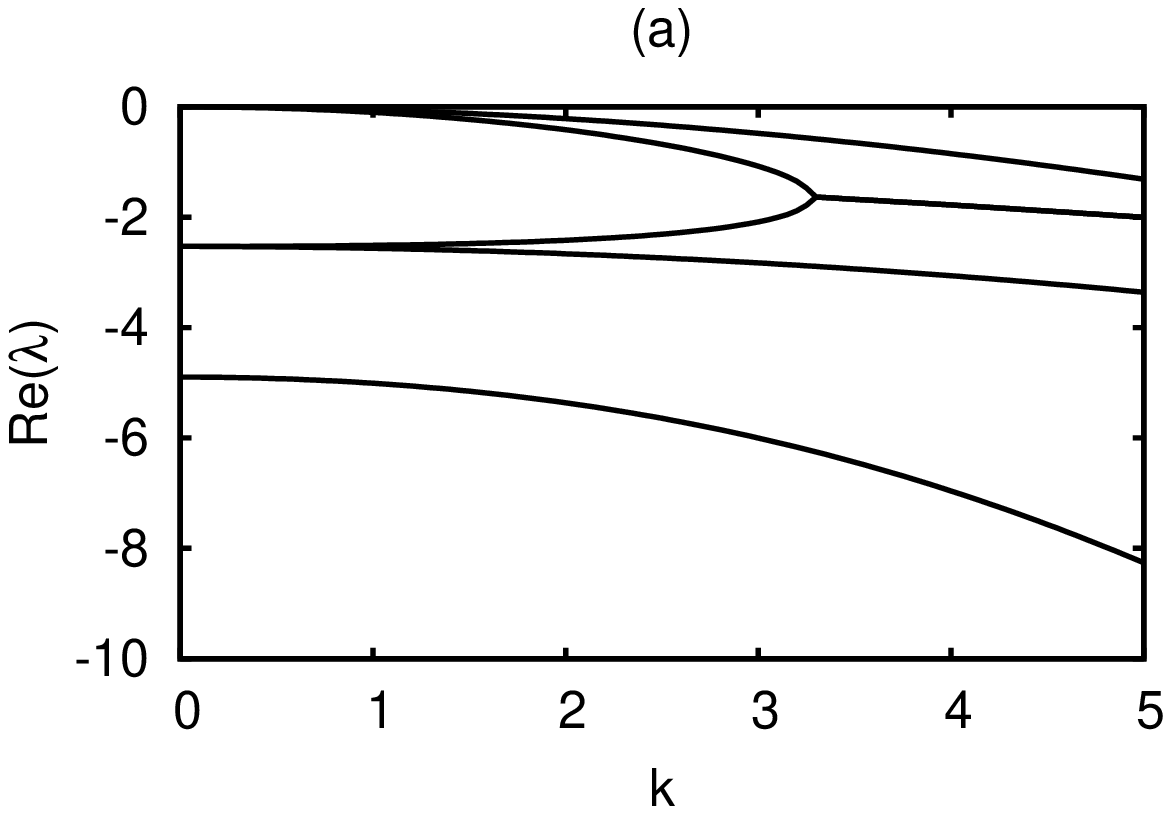}
  \includegraphics[width=.45\textwidth]{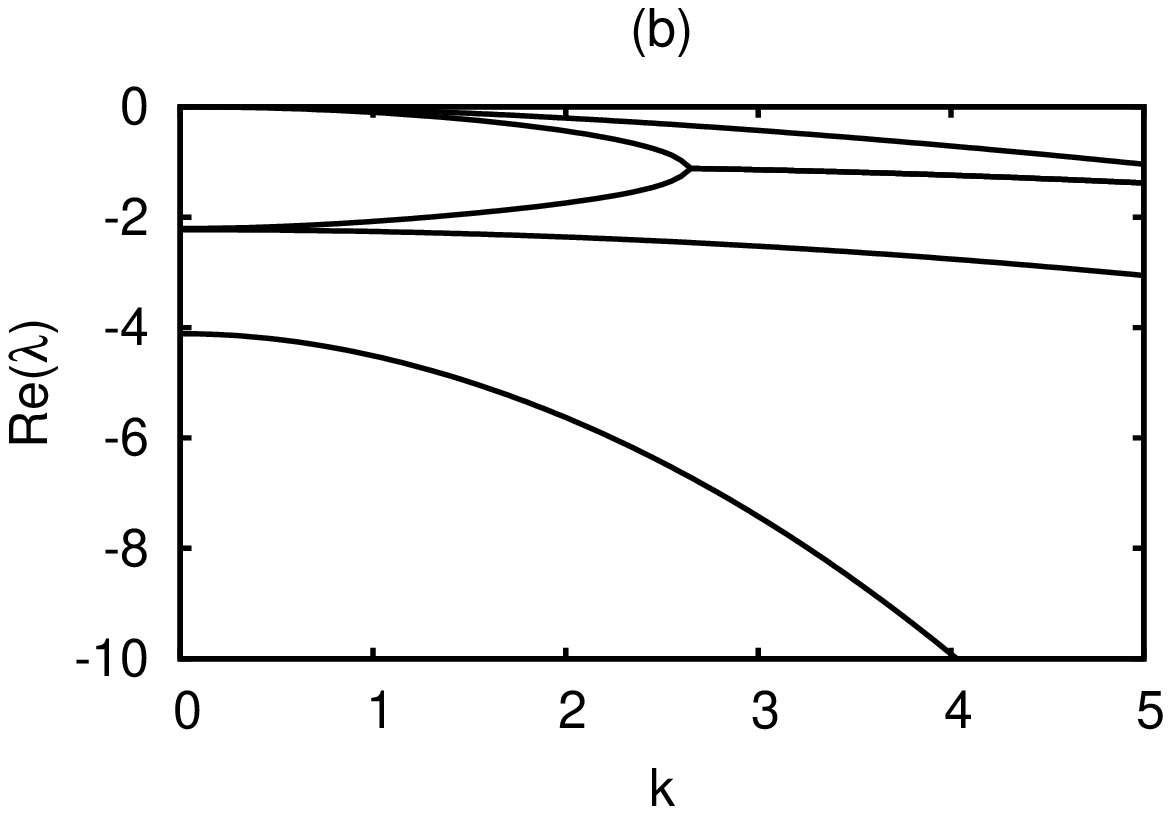}
  \includegraphics[width=.45\textwidth]{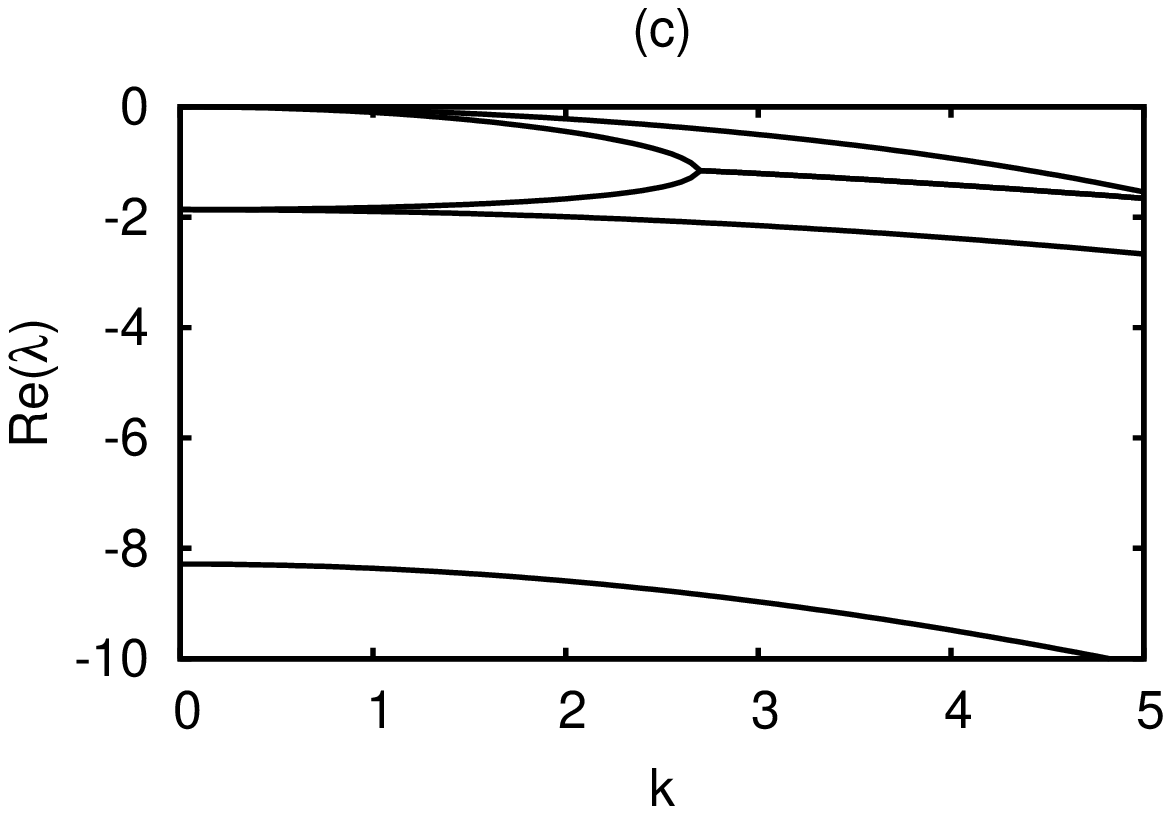}
  \caption{Real part of the transversal and longitudinal eigenvalues for $\alpha=0.9$, $d=3$, $\sigma_1=\sigma_2$, $x_1=0.2$, and three different values of the mass ratio:  $m_1/m_2=2$  (panel a), $m_1/m_2=3$ (panel b), and $m_1/m_2=4$ (panel c). The parameters of the driven model are the same as those considered in Figs.\ \ref{fig:1}--\ref{fig:4}}
  \label{fig:5}
\end{figure}

\subsection{Euler hydrodynamics ($k=0$). Some special cases}

For an inviscid fluid ($k=0$), the square matrix $\mathsf{M}$ reduces to the matrix $\mathsf{M}^{(0)}$. Even in this limit case, the eigenvalues of $\mathsf{M}^{(0)}$ must be numerically obtained by solving the quartic equation \eqref{4.20}. On the other hand, these eigenvalues can be analytically determined for some special systems.

In particular, for mechanically equivalent particles, $D_U^*=0$,  $\chi_i=1$, and the eigenvalues of the longitudinal hydrodynamic modes are
\beq
\label{4.22.1}
\lambda_{\parallel}=\left(0,0,-\frac{3}{2}\zeta^*-2^{-(\beta-1)}\gamma^*,-2^{-\beta}\gamma^*\right),
\eeq
where use has been made of the condition \eqref{4.0} for the present case. Consequently, since the eigenvalues are zero or negative, the HSS is linearly stable. For $k \neq 0$, it is easy to see that the critical value $k_{\parallel}^c$ (defined as $\max\lambda_\parallel(k_\parallel^c)=0$) is negative and hence, the HSS is again always stable. This conclusion agrees with the previous stability analysis carried out for single granular gases \cite{gachve13}.

Let us consider now elastic collisions ($\al_{ij}=1$). In this case, $\zeta_i^*=0$, $\chi_i=1$  and in the absence of spatial gradients ($k=0$) the eigenvalues are
\beq
\label{4.24}
\lambda_{\parallel}=\left(0,0,-2\gamma^*(\mu_{12}^\beta+x_1 \delta m_\beta),\lambda_\perp\right).
\eeq
Since $(\mu_{12}^\beta+x_1 \delta m_\beta)>0$ and $\lambda_\perp<0$ (see Eq.\ \eqref{4.8} for $k=0$), then $\lambda_{\parallel} \leq 0$.

\subsection{General case}

For general binary mixtures, the wave vector dependence of the four longitudinal hydrodynamic modes is more complex. This can be achieved by numerically solving Eq.\ \eqref{4.20}. As an illustration, Fig.\ \ref{fig:5} shows the real parts of the transversal and longitudinal modes $\lambda(k)$ for the (common) coefficient of restitution $\al=0.9$, $d=3$, $\sigma_1=\sigma_2$, $x_1=0.2$, the mass ratios $m_1/m_2=2$, 3, and 4, and for the same driven parameters as in Figs.\ \ref{fig:1}--\ref{fig:4}. We observe that the six hydrodynamic modes have two different degeneracies. As occurs for undriven granular mixtures \cite{GMD06}, while the (transversal) shear mode degeneracy remains at finite $k$ the other degeneracy is removed at any finite value of the wave number $k$. In particular, two real modes become a conjugate complex pair for wave numbers larger than a certain value. In any case, it is quite apparent that $\text{Re}(\lambda)\leq 0$ and hence the HSS is (linearly) stable in the complete range of wave numbers studied.

In general, one of the longitudinal modes could be unstable for $k<k_{\parallel}^c$ where the critical longitudinal mode $k_{\parallel}^c$ can be obtained from Eq.\ \eqref{4.20} when $\lambda=0$. A careful analysis leads to the equation
\beq
\label{4.21}
\det(M)=k^4(X_2 \; +X_4 \; k^2)=0,
\eeq
where the coefficients $X_2$ and $X_4$ are known functions of the parameters of the problem. Their explicit forms are very large and will be omitted here. The solutions to Eq.\ \eqref{4.21} lead to the critical values
\beq
\label{4.24}
k_{\parallel}^c=\left(0,0,0,0,-\sqrt{-\frac{X_2}{X_4}},\sqrt{-\frac{X_2}{X_4}}\right).
\eeq
A systematic analysis of the dependence of the sign of $X_2/X_4$ on the control parameters shows that this ratio is always positive. This means that there are no physical values of the wave numbers for which the longitudinal modes become unstable. Therefore, as in the case of the transversal shear mode, we can conclude that all the eigenvalues of the dynamical matrix $\mathsf{M}$ have a negative real part and no instabilities are found for driven granular mixtures.

\section{Discussion}
\label{sec5}

In this paper, the Navier--Stokes hydrodynamic equations for a low-density driven granular binary mixture have been discussed. The mixture is driven by a stochastic bath with friction. The form of the fluxes of mass, momentum, and energy have been derived and the corresponding set of transport coefficients identified. The associated transport coefficients are the diffusion coefficient $D$, the pressure diffusion coefficient $D_p$, the thermal diffusion coefficient $D_T$, and the velocity diffusion coefficient $D_U$ in the case of the mass flux, the shear viscosity coefficient $\eta$ for the pressure tensor, the Dufour coefficient $D''$, the pressure energy coefficient $L$, the thermal conductivity $\kappa$, and the velocity conductivity $\kappa_U$ in the case of the heat flux. As occurs for undriven granular mixtures \cite{GD02,GMD06,GDH07,GHD07}, the above transport coefficients are determined from the solutions of a set of coupled linear integral equations. On the other hand, for practical purposes, these integral equations are usually solved by considering the leading terms in Sonine polynomial expansions. Although these truncations are expected to be unreliable at extreme values of mass and size ratio \cite{MG03,GM04,GV09,GV12}, they can be still considered as accurate for many different situations of practical interest.

Since the dependence of the diffusion ($D$,$D_p$, $D_T$, and $D_U$) and shear viscosity $\eta$ coefficients on the control parameters was widely studied in a previous paper \cite{KG13}, attention has been focused here on the remaining four transport coefficients associated with the heat flux. In this case, the second Sonine approximation has been considered to provide the explicit dependence of the set $\left\{D'',L,\kappa,\kappa_U\right\}$ on the mass and size ratios, composition, coefficients of restitution, and the driven parameters of the model. As has been remarked in previous papers \cite{GD02,GMD06,GDH07,GHD07}, there is no phenomenology involved since the transport coefficients have been derived systematically by solving the (inelastic) Boltzmann kinetic equation from the Chapman--Enskog method. Therefore, there is no \emph{a priori} limitation on the degree of inelasticity as the transport coefficients are highly nonlinear functions of the coefficients of restitution. In addition, in contrast to previous results for granular mixtures \cite{JM89,Z95,AW98,WA99,SGNT06}, the influence of the nonequipartition of energy on transport has been accounted for and the Navier--Stokes transports coefficients depend on the temperature ratios $T_i/T$ and their derivatives with respect to both the composition and the driven parameters of the model \cite{KG13}. These latter derivatives introduce conceptual and practical difficulties not present in the case of undriven granular mixtures \cite{GD02,GMD06,GDH07,GHD07}.

In the same way as for the diffusion and shear viscosity coefficients \cite{KG13}, Figs.\ \ref{fig:1}--\ref{fig:4} highlight the impact of inelasticity on the heat flux transport coefficients since their forms are clearly different from those obtained for elastic collisions. This is specially relevant in the case of the (reduced) transport coefficient $L_p^*$ since it vanishes for elastic collisions. In addition, depending on the value of the mass ratio, in most of the cases the (reduced) heat flux transport coefficients monotonically increase or decrease with inelasticity. An exception is the (reduced) thermal diffusion factor $\kappa_T$ which exhibits a non-monotonic dependence on the coefficient of restitution when the mass of the defect component is lighter than that of the excess component. With respect to the influence of thermostats, it is seen that they play an important role in the transport of energy since the behavior of the heat flux transport coefficients differs from the one found for undriven granular mixtures \cite{GMD06}.

As an application of the previous results, the stability of the special HSS solution has been analyzed. This has been achieved by solving the linearized Navier--Stokes hydrodynamic equations for small perturbations around the HSS. The linear stability analysis performed here show no new surprises relative to the earlier work carried out for monocomponent driven granular gases \cite{gachve13}: the HSS is linearly stable with respect to long enough wavelength excitations. The only difference with respect to the single case is the addition of the stable mass diffusion mode. Of course, the quantitative features can be quite different as there are additional degrees of freedom with the parameter set $\left\{x_1,T, m_1/m_2,\sigma_1/\sigma_2,\alpha_{ij}\right\}$. The conclusion reached here for the reference HSS differs from the one found for freely cooling granular mixtures where it was shown that the resulting hydrodynamic equations exhibit a long wavelength instability for three of the hydrodynamic modes. This shows again the influence of thermostats on the dynamics of granular flows.

It is quite apparent that the theoretical results obtained in this paper for the stability of the HSS should be tested against computer simulations. This would allow us to gauge the degree of accuracy of the theoretical predictions. As happens for undriven granular gases \cite{BRM98,MDCPH11,MGHEH12,BR13,MGH14}, we expect that the present results stimulate the performance of appropriate simulations where our theory can be assessed. We also plan to undertake such kind of simulations in the near future. Finally, as alluded to in Ref.\ \cite{KG13}, we think that our results can be also relevant for practical purposes since most of the simulations reported in granular literature \cite{puglisi,GSVP11,ernst} have been performed by using external driving forces. Given that in many of the above papers the \emph{elastic} forms of the transport coefficients have been employed to compare simulations and theory, it is quite evident that the results provided here can be useful for simulators interested in both driven granular mixtures or bidisperse granular suspensions.

\acknowledgments

The research of N.K. and V.G. has been supported by the Ministerio de Econom\'ia y Competitividad (Spain) through grants FIS2015-63628-C2-1-R and FIS2016-76359-P, respectively, both partially financed by ``Fondo Europeo de Desarrollo Regional'' funds. The research of VG has also been supported by the Junta de Extremadura (Spain) through Grant No. GR15104, partially financed by ``Fondo Europeo de Desarrollo Regional'' funds.

\appendix

\section{Explicit calculations of the heat flux transport coefficients}
\label{appA}

In this Appendix, we provide some technical results for the determination of the transport coefficients associated with the heat flux. These coefficients are defined as \cite{KG13}
\begin{equation}
\label{a1}
D''=-\frac{1}{dT^{2}}\sum_{i=1}^2\,\int d\mathbf{ v}\,\frac{1}{2}
m_{i}V^{2}\mathbf{V}\cdot {\boldsymbol{\cal A}}_{i},
\end{equation}
\begin{equation}
\label{a2}
L=-\frac{1}{d}\sum_{i=1}^2\,\int d\mathbf{ v}\,\frac{1}{2}m_{i}V^{2}\mathbf{V}
\cdot {\boldsymbol{\cal B}}_{i},
\end{equation}
\begin{equation}
\label{a3}
\kappa =-\frac{1}{d}\sum_{i=1}^2\,\int d\mathbf{ v}\,\frac{1}{2}m_{i}V^{2}
\mathbf{V}\cdot {\boldsymbol{\cal C}}_{i},
\end{equation}
\begin{equation}
\label{a4}
\kappa_U =-\frac{1}{d}\sum_{i=1}^2\,\int d\mathbf{ v}\,\frac{1}{2}m_{i}V^{2}
\mathbf{V}\cdot {\boldsymbol{\cal G}}_{i}.
\end{equation}
The quantities $\left\{{\boldsymbol{\cal A}}_{i}, {\boldsymbol{\cal B}}_{i}, {\boldsymbol{\cal C}}_{i}\right\}$ and ${\boldsymbol{\cal G}}_{i}$ obey Eqs.\ (64)--(66) and (69), respectively, of Ref.\ \cite{KG13}.

The evaluation of these transport coefficients requires going up to the second Sonine approximation. In this case, the quantities ${\boldsymbol{\cal A}}_{i}$, ${\boldsymbol{\cal B}}_{i}$, ${\boldsymbol{\cal C}}_{i}$, and ${\boldsymbol{\cal G}}_{i}$ are given by
\begin{widetext}
\beq
\label{a5}
{\boldsymbol{\cal A}}_{1} \to f_{1,\text{M}}\left[ -\frac{m_1 m_2 n}{\rho n_1 T_1}D \mathbf{V}+d_1'' \mathbf{S}_1(\mathbf{V})\right], \quad
{\boldsymbol{\cal A}}_{2} \to f_{1,\text{M}}\left[ \frac{m_1 m_2 n}{\rho n_2 T_2}D \mathbf{V}+d_2'' \mathbf{S}_2(\mathbf{V})\right],
\eeq
\beq
\label{a6}
{\boldsymbol{\cal B}}_{1} \to f_{1,\text{M}}\left[ -\frac{\rho}{p n_1 T_1}D_p \mathbf{V}+\ell_1 \mathbf{S}_1(\mathbf{V})\right],
\quad {\boldsymbol{\cal B}}_{2} \to f_{2,\text{M}}\left[ -\frac{\rho}{p n_2 T_2}D_p \mathbf{V}+\ell_2 \mathbf{S}_2(\mathbf{V})\right],
\eeq
\beq
\label{a7}
{\boldsymbol{\cal C}}_{1} \to f_{1,\text{M}}\left[ -\frac{\rho}{T n_1 T_1}D' \mathbf{V}+\kappa_1 \mathbf{S}_1(\mathbf{V})\right],
\quad {\boldsymbol{\cal C}}_{2} \to f_{2,\text{M}}\left[ -\frac{\rho}{T n_2 T_2}D' \mathbf{V}+\kappa_2 \mathbf{S}_2(\mathbf{V})\right],
\eeq
\beq
\label{a8}
{\boldsymbol{\cal G}}_{1} \to f_{1,\text{M}}\left[ -\frac{1}{n_1 T_1}D_U \mathbf{V}+\kappa_{U1} \mathbf{S}_1(\mathbf{V})\right],
\quad {\boldsymbol{\cal G}}_{2} \to f_{2,\text{M}}\left[\frac{1}{n_2 T_2}D_U \mathbf{V}+\kappa_{U2} \mathbf{S}_2(\mathbf{V})\right],
\eeq
\end{widetext}
where
\beq
\label{a9}
\mathbf{S}_i(\mathbf{V})=\left(\frac{m_i}{2}V^2-\frac{d+2}{2}T_i\right)\mathbf{V}.
\eeq
In Eqs.\ \eqref{a5}--\eqref{a8}, it is understood that $D$, $D_p$, $D'$, and $D_U$ are obtained in the first Sonine approximation, namely, they are given by Eqs.\ (86)--(89) of Ref.\ \cite{KG13}. The coefficients $d_i''$, $\ell_i$, $\kappa_i$, and $\kappa_{Ui}$ are defined as
\beq
\label{a10}
\left(
\begin{array}{c}
d_i''\\
\ell_i\\
\kappa_i\\
\kappa_{Ui}
\end{array}
\right)=\frac{2}{d(d+2)}\frac{m_i}{n_iT_i^3}\int \dd \mathbf{v}\;\mathbf{S}_i(\mathbf{V})\cdot
\left(
\begin{array}{c}
{\boldsymbol{\cal A}}_{i}\\
{\boldsymbol{\cal B}}_{i}\\
{\boldsymbol{\cal C}}_{i}\\
{\boldsymbol{\cal G}}_{i}
\end{array}
\right).
\eeq
The coefficients $d_i''$, $\ell_i$, $\kappa_i$ can be determined by multiplying Eqs.\ (64)--(66) (and their counterparts for species $2$) of Ref.\ \cite{KG13} by $\mathbf{S}_i (\mathbf{V})$ and integrating over velocity. Analogously, the coefficients $\kappa_{Ui}$ are obtained from Eq.\ (69) of Ref.\ \cite{KG13} by following similar steps. The final expressions can be obtained by taking into account the results
\beq
\label{a11}
\int\; \dd\mathbf{v}\; m_i \mathbf{S}_i\cdot \mathbf{A}_i=-\frac{d(d+2)}{2}\frac{n_iT_i T}{m_i}\frac{\partial \chi_i}{\partial x_1},
\eeq
\beq
\label{a12}
\int\; \dd\mathbf{v}\; m_i \mathbf{S}_i\cdot \mathbf{B}_i=\frac{d(d+2)}{2}\frac{n_iT_i T}{pm_i}\left( \xi^* \frac{\partial \chi_i}{\partial \xi^*}+\frac{2}{3} \omega^* \frac{\partial \chi_i}{\partial \omega^*}\right),
\eeq
\beqa
\label{a13}
\int\; \dd\mathbf{v}\; m_i \mathbf{S}_i\cdot \mathbf{C}_i&=&-\frac{d(d+2)}{2}\frac{n_iT_i }{m_i}\left(
\chi_i -\frac{1}{2} \xi^* \frac{\partial \chi_i}{\partial \xi^*}\right. \nonumber\\
& & \left.+\frac{2}{3} \omega^* \frac{\partial \chi_i}{\partial \omega^*}
\right),
\eeqa
\beq
\label{a14}
\int\; \dd\mathbf{v}\; m_i \mathbf{S}_i\cdot \mathbf{G}_i=0,
\eeq
where $\mathbf{A}_i$, $\mathbf{B}_i$, and $\mathbf{C}_i$ are defined by Eqs.\ (B10)--(B12) of Ref.\ \cite{KG13} and the derivatives of the temperature ratio $\chi_i$ with respect to $x_1$, $\omega^*$, and $\xi^*$ have been also determined in the Appendix A of Ref.\ \cite{KG13}. In Eqs.\ \eqref{a12} and \eqref{a13}, we have introduced the dimensionless quantities
\beq
\label{a15}
\omega^*\equiv \frac{\gamma_\text{b}}{\overline{m}^\beta} \left(\frac{\overline{m}^\lambda}{2 \xi_\text{b}^2}\right)^{1/3} \left(n \sigma_{12}^{d-1}\right)^{-2/3},
\eeq
\beq
\label{a16}
\xi^* \equiv \frac{\xi_\text{b}^2}{n \sigma_{12}^{d-1}\overline{m}^{\lambda-1} T v_0}.
\eeq

The set of algebraic equations obeying the reduced coefficients $\kappa_{Ui}^*\equiv (\rho/p T)\kappa_{Ui}$ is decoupled from the remaining coefficients. In matrix form, the coefficients $\kappa_{U1}^*$ and $\kappa_{U2}^*$ are obtained by solving the set of linear equations
\beq
\label{a17}
\left(
\begin{array}{cc}
b_{11}& b_{12} \\
b_{21}& b_{22}
\end{array}
\right)
\left(
\begin{array}{c}
\kappa_{U1}^* \\
\kappa_{U2}^*
\end{array}
\right)
= \left(
\begin{array}{c}
c_7 \\
c_8
\end{array}
\right),
\end{equation}
where
\beq
\label{a18}
b_{11}=\chi_1^3 \left(3\frac{\omega^* \xi^{*1/3}}{M_1^\beta}+\nu_{11}\right),
\eeq
\beq
\label{a19}
b_{12}= \nu_{12} \chi_1^3, \quad b_{21}=\nu_{21}\chi_2^3,
\eeq
\beq
\label{a20}
b_{22}=\chi_2^3 \left(3\frac{\omega^* \xi^{*1/3}}{M_2^\beta}+\nu_{22}\right),
\eeq
\beq
\label{a21}
c_7= \frac{n\overline m}{2\rho}e_{12} D_U^*, \quad c_8= -\frac{n\overline m}{2\rho}e_{21} D_U^*.
\eeq
Here, we recall that $M_i\equiv m_i/\overline{m}$ and
\begin{equation}
\label{a22}
e_{ij}=\frac{1}{x_i}\Bigg[-\frac{\xi^*}{M_i^{\lambda-1}} +\chi_i\left(2\frac{\omega^* \xi^{*1/3}}{M_i^\beta}+\omega_{ij}\right)\Bigg].
\end{equation}
The explicit forms of the (reduced) collision frequencies $\omega_{ij}$, $\nu_{ii}$, and $\nu_{ij}$ are given by Eqs.\ (9.16)--(9.18), respectively, of Ref.\ \cite{gamo07}. The solution to Eq.\ \eqref{a17} is simply given by
\beq
\label{a23}
\kappa_{U1}^*=\frac{b_{22}c_7-b_{12}c_8}{b_{11}b_{22}-b_{12}b_{21}}, \quad
\kappa_{U2}^*=\frac{b_{11}c_8-b_{21}c_7}{b_{11}b_{22}-b_{12}b_{21}}.
\eeq

We consider now the coefficients $d_i''$, $\ell_i$, and $\kappa_i$. By using matrix notation, the coupled set of six algebraic equations for the reduced coefficients
\beq
\label{a24}
\left\{d_1^*,d_2^*,\ell_1^*, \ell_2^*, \kappa_1^*, \kappa_2^*\right\}
\eeq
can be written as
\beq
\label{a25}
\Lambda_{\mu \mu'}X_{\mu'}=Y_\mu.
\eeq
Here, $d_i^*\equiv T\nu_0 d_i''$, $\ell_i^*\equiv p T \nu_0 \ell_i$, and $\kappa_i^* \equiv T^2 \nu_0 \kappa_i$, where $\nu_0$ is defined by Eq.\ \eqref{nu0}. In addition,  $X_{\mu'}$ is the column matrix defined by the set \eqref{a24}, $\Lambda_{\mu \mu'}$ is the square matrix
\beq
\label{a26}
\boldsymbol{\Lambda}=
\left(
\begin{array}{cccccc}
b_{11}& b_{12} & b_{13}& 0 & b_{13} & 0 \\
b_{21}& b_{22} & 0 & b_{24}& 0 & b_{24}  \\
0&0& b_{33} & b_{12} & b_{35} &0 \\
0&0& b_{21} & b_{44} & 0 &b_{46} \\
0&0& b_{53} & 0 & b_{55} &b_{12} \\
0&0& 0 & b_{64} & b_{21} &b_{66} \\
\end{array}
\right),
\eeq
and
\beq
\label{a27}
Y=\left(
\begin{array}{c}
c_1 \\ c_2 \\ c_3 \\ c_4 \\ c_5 \\ c_6
\end{array}
\right).
\eeq
Here, we have introduced the (dimensionless) quantities
\beq
\label{a28}
b_{13}=\chi_1^3 \bigg(\xi^* \delta m_{\lambda-1} -2\omega^*\xi^{*1/3} \delta m_{\beta}\frac{\partial (x_1 \chi_1)}{\partial x_1}-\frac{1}{\nu_0}\frac{\partial \zeta^{(0)}}{\partial x_1} \bigg),
\eeq
\beq
\label{a29}
b_{24}=\left(\frac{\chi_2}{\chi_1}\right)^3 b_{13},
\eeq
\beqa
\label{a30}
b_{33}&=&-\chi_1^3\bigg[2\omega^* \xi^{*1/3}\bigg( \sum_{i=1}^2 \frac{x_i\chi_i}{M_i^\beta}
    +\delta m_\beta x_1 p\frac{\partial \chi_1}{\partial p} \bigg)\nonumber\\
& & -\xi^*  \sum_{i=1}^2 \frac{x_i}{M_i^{\lambda-1}}
    +\frac{\zeta^{(0)}}{\nu_0}+\frac{p}{\nu_0}\frac{\partial \zeta^{(0)}}{\partial p} \bigg] +b_{11},\nonumber\\
\eeqa
\beq
\label{a31}
b_{35}=-\chi_1^3\left(2\omega^* \xi^{*1/3} \delta m_\beta x_1 p\frac{\partial \chi_1}{\partial p} +\frac{p}{\nu_0}\frac{\partial \zeta^{(0)}}{\partial p} \right),
\eeq
\beqa
\label{a32}
b_{44}&=&-\chi_2^3\bigg[2\omega^* \xi^{*1/3}\bigg( \sum_{i=1}^2 \frac{x_i\chi_i}{M_i^\beta}
    +\delta m_\beta x_1 p\frac{\partial \chi_1}{\partial p} \bigg)\nonumber\\
& & -\xi^*  \sum_{i=1}^2 \frac{x_i}{M_i^{\lambda-1}}
    +\frac{\zeta^{(0)}}{\nu_0}+\frac{p}{\nu_0}\frac{\partial \zeta^{(0)}}{\partial p} \bigg]+b_{22}, \nonumber\\
\eeqa
\beq
\label{a33}
b_{46}=\left(\frac{\chi_2}{\chi_1}\right)^3 b_{35},
\eeq
\beqa
\label{a34}
b_{53}&=&-\chi_1^3\bigg(2\omega^* \xi^{*1/3} \delta m_\beta x_1 T\frac{\partial \chi_1}{\partial T} \nonumber\\
& & + \xi^*  \sum_{i=1}^2 \frac{x_i}{M_i^{\lambda-1}} +\frac{T}{\nu_0}\frac{\partial \zeta^{(0)}}{\partial T} \bigg),
\eeqa
\beqa
\label{a35}
b_{55}&=&-\chi_1^3 \bigg[2\omega^* \xi^{*1/3}\bigg( \sum_{i=1}^2 \frac{x_i\chi_i}{M_i^\beta}
    +\delta m_\beta x_1 T\frac{\partial \chi_1}{\partial T} \bigg)\nonumber\\
    & & +\frac{\zeta^{(0)}}{\nu_0}+\frac{T}{\nu_0}\frac{\partial \zeta^{(0)}}{\partial T} \bigg] +b_{11},
\eeqa
\beq
\label{a36}
b_{64}=\left(\frac{\chi_2}{\chi_1}\right)^3 b_{53},
\eeq
\beqa
\label{a37}
b_{66}&=&-\chi_2^3 \bigg[2\omega^* \xi^{*1/3}\bigg( \sum_{i=1}^2 \frac{x_i\chi_i}{M_i^\beta}
    +\delta m_\beta x_1 T\frac{\partial \chi_1}{\partial T} \bigg)\nonumber\\
    & & +\frac{\zeta^{(0)}}{\nu_0}+\frac{T}{\nu_0}\frac{\partial \zeta^{(0)}}{\partial T} \bigg] +b_{22}.
\eeqa
The coefficients $c_1$--$c_6$ are defined as
\beq
\label{a38}
c_1=-\chi_1 \frac{\partial \chi_1}{\partial x_1}+e_{12} D^*, \quad
c_2=-\chi_2 \frac{\partial \chi_2}{\partial x_1}-e_{21}D^*,
\eeq
\beq
\label{a39}
c_3=\frac{\chi_1}{3}\left(3\xi^*\frac{\partial \chi_1}{\partial \xi^*}+2\omega^* \frac{\partial \chi_1}{\partial \omega^*} \right)
+e_{12} D_p^*,
\eeq
\beq
\label{a40}
c_4=\frac{\chi_2}{3}\left(3\xi^*\frac{\partial \chi_2}{\partial \xi^*}+2\omega^* \frac{\partial \chi_2}{\partial \omega^*} \right)
-e_{21} D_p^*,
\eeq
\beq
\label{a41}
c_5=\frac{\chi_1}{6}\left(3\xi^*\frac{\partial \chi_1}{\partial \xi^*}-6\chi_1-4\omega^* \frac{\partial \chi_1}{\partial \omega^*} \right)+e_{12} D_T^*,
\eeq
\beq
\label{a42}
c_6=\frac{\chi_2}{6}\left(3\xi^*\frac{\partial \chi_2}{\partial \xi^*}-6\chi_2-4\omega^* \frac{\partial \chi_2}{\partial \omega^*} \right)-e_{21} D_T^*.
\eeq
In the above equations, $\zeta^{(0)}$ is the zeroth-order contribution to the cooling rate and $\delta m_\beta$ is defined by Eq.\ \eqref{a43}. The solution to Eq.\ \eqref{a25} is
\beq
\label{a44}
X_\mu=\left(\Lambda^{-1}\right)_{\mu\mu'}Y_{\mu'}.
\eeq
From this relation one gets the expressions of the coefficients $d_i^*$, $\ell_i^*$, and $\kappa_i^*$ in terms of the parameters of the mixture.

\end{document}